# Conditional Segmentation in Lieu of Image Registration

Yipeng Hu[1,2], Eli Gibson[3], Dean Barratt[1], Mark Emberton[4], Alison Noble[2], Tom Vercauteren[5]

[1] Centre for Medical Image Computing and Wellcome/EPSRC Centre for Interventional & Surgical Sciences, University College London, London, UK
[2] Institute of Biomedical Engineering, University of Oxford, Oxford, UK
[3] Digital Services, Digital Technology & Innovation, Siemens Healthineers, Princeton, NJ, USA
[4] Division of Surgery & Interventional Science, University College London, London, UK
[5] School of Biomedical Engineering & Imaging Sciences, King's College London, London, UK
yipeng.hu@ucl.ac.uk

**Abstract.** Classical pairwise image registration methods search for a spatial transformation that optimises a numerical measure that indicates how well a pair of moving and fixed images are aligned. Current learning-based registration methods have adopted the same paradigm and typically predict, for any new input image pair, dense correspondences in the form of a dense displacement field or parameters of a spatial transformation model. However, in many applications of registration, the spatial transformation itself is only required to propagate points or regions of interest (ROIs). In such cases, detailed pixel- or voxel-level correspondence within or outside of these ROIs often have little clinical value. In this paper, we propose an alternative paradigm in which the location of corresponding image-specific ROIs, defined in one image, within another image is learnt. This results in replacing image registration by a *conditional segmentation* algorithm, which can build on typical image segmentation networks and their widely-adopted training strategies. Using the registration of 3D MRI and ultrasound images of the prostate as an example to demonstrate this new approach, we report a median target registration error (TRE) of 2.1 mm between the ground-truth ROIs defined on intraoperative ultrasound images and those propagated from the pre-operative MR images. Significantly lower (>34%) TREs were obtained using the proposed conditional segmentation compared with those obtained from a previously-proposed spatial-transformation-predicting registration network trained with the same multiple ROI labels for individual image pairs. We conclude this work by using a quantitative bias-variance analysis to provide one explanation of the observed improvement in registration accuracy.

## 1   Introduction

Recent medical image registration methods based on convolutional neural networks have adopted an end-to-end learning framework, in which a *moving* and *fixed* image pair is the input of the network that directly predicts a dense displacement field (DDF) or parameters of a parametric spatial transformation model [1]. These capture pixel- or voxel-level *dense correspondences*. Such networks have been trained by minimising unsupervised losses [2], adapted from classical or learned dissimilarity measures within pairs of images, or supervised losses measuring the difference to ground-truth transformations [3, 4]. Label similarity between anatomical segmentations has also been proposed to measure image alignment as a form of weak supervision [5] and has been combined with other losses [6, 7].

Predicting spatial transformations, as in the above-mentioned methods, enables physically-motivated prior knowledge on the deformation fields to be incorporated in the network training. Examples include parameterising the spatial transformation using rigid, spline-based models [1-3] or velocity fields [4], penalising implausible transformation through a regularisation term such as $L^2$-norm of DDF and bending energy [5-7], and minimising divergence between the predicted and the unpaired ground-truth deformations [8]. Fig. 1 illustrates the training and prediction stages of these typical spatial-transformation-predicting registration networks.

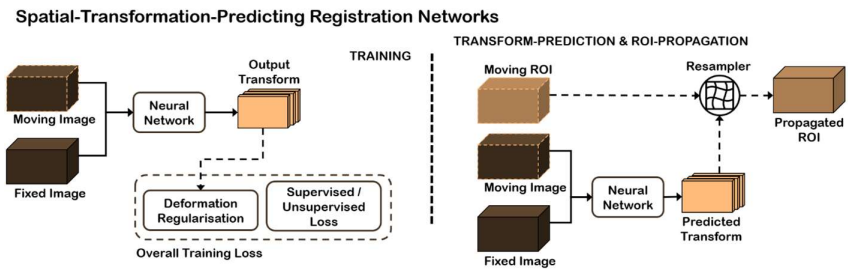

**Fig. 1.** Left: illustration of the training of spatial-transformation-predicting registration networks; Right: the ROIs are propagated from moving image to fixed image by the spatial transformation (predicted transform), predicted by the trained neural network.

A common purpose of image registration in medical applications is region-of-interest (ROI) propagation, also illustrated in Fig. 1. In multimodal image guided interventions, for instance, the registration-generated spatial transformation is used to warp one or more clinically useful ROIs, defined by inseparable pixel/voxel locations in the preoperative moving images, to intraoperative fixed images. These ROIs, such as patient-specific biopsy or pathology locations in a preoperative-image-derived procedure plan, are not necessarily anatomically-defined landmarks or identifiable in both images, and therefore are not consistently available for all cases *a priori*. Different to segmentation, a registration network predicts spatial transformation that defines dense correspondence, which can propagate any given ROI from a moving image to a fixed image.

However, clinically useful ROIs are often sparse for individual patients (e.g. a single target tumour), therefore propagating ROIs (equivalent to searching for a region-level correspondence) and searching for dense correspondence present very different challenges. This is partly because the regularised dense correspondence prediction encourages spatial smoothness and topology preservation, which may over-constrain localising ROIs in the fixed images that is of much greater clinical value. We postulate that, with the increasing availability in training data, predicting spatial transformation may limit the *clinically relevant registration accuracy* in such applications.

In this work, we propose to use a machine learning approach for *any* given ROI on a moving image to predict the image-specific ROI on a fixed image directly. This approach does not predict a spatial transformation and does not require deformation regularisation. Replacing the task of finding a spatial transformation with ROI propagation leads to a *conditional segmentation* approach, which is described in Section 2.

We report experimental results from a multimodal image registration application, in which MR and ultrasound images are aligned to guide targeted biopsy [9] and focal therapy [10] for prostate cancer patients. The contributions of this work include: 1) a novel conditional segmentation paradigm for ROI propagation tasks, which replaces commonly-adopted image registration methods; 2) the demonstration using clinical data that significantly improved registration accuracy can be achieved using the proposed conditional segmentation approach compared with a spatial-transformation-predicting registration network for a real-world application; and 3) a bias-variance analysis to further investigate the source of improvement shown for this application.

## 2 Method

### 2.1 Conditional Image Segmentation Paradigm for ROI Propagation

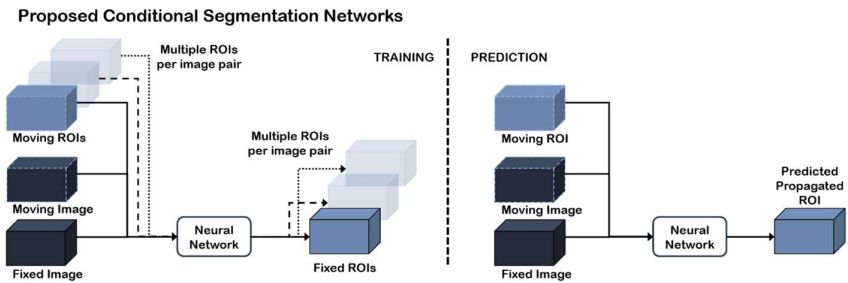

**Fig. 2.** Left: illustration of the training of the proposed conditional segmentation network; Right: individually propagated ROIs are directly predicted by the trained neural network.

With clinically relevant ROIs having varying quantities, shapes, sizes and locations for each image pair, we formulate the ROI propagation task as a joint binary classification problem where each voxel on the fixed image is to be classified as either "belonging to" ($C_{k=1}$) or "not belonging to" ($C_{k=2}$) the ROI propagated from the moving image. Using a convolutional neural network with parameters $\boldsymbol{\theta}$, the posterior class probabilities, modelled by the network output, are given by $p_{\boldsymbol{\theta}}(C_k|\mathbf{I}^{fix}, \mathbf{I}^{mov}, \mathbf{R}^{mov})$, where random vectors $\mathbf{I}^{fix}$, $\mathbf{I}^{mov}$ and $\mathbf{R}^{mov}$ represent the fixed image, the moving image and the ROI in the moving image (hereafter referred to as the "moving ROI"), respectively. Predicting $p_{\boldsymbol{\theta}}(C_k|\mathbf{I}^{fix}, \mathbf{I}^{mov}, \mathbf{R}^{mov})$, in turn, represents a conditional segmentation problem, conditioned on a given moving image $\mathbf{I}^{mov}$ and a given moving ROI $\mathbf{R}^{mov}$.

As illustrated in Fig. 2, the conditional segmentation can be implemented with minimal adaptation to a standard image segmentation network by, for instance, concatenating the image pair ($\mathbf{I}^{fix}, \mathbf{I}^{mov}$) and moving ROI $\mathbf{R}^{mov}$ in the input layer. Unlike conventional spatial-transformation-predicting registration or multi-ROI segmentation methods, this network predicts any one single propagated ROI $\mathbf{R}^{fix}$ (potentially with single foreground voxel) at a time during inference stage and can be trained with multiple training ROIs labelled from each of the training image pairs.

### 2.2 A Supervised Training Approach

In this work, we implement a supervised conditional segmentation training approach, suitable for the multimodality 3D image registration application described in Section 3.

In training, $N$ pairs of moving images $\{\mathbf{i}_n^{mov}\}$ and fixed images $\{\mathbf{i}_n^{fix}\}$ are available, $n = 1, \ldots, N$. For every $n^{th}$ image pair, $M_n$ pairs of corresponding ROI labels $\{\mathbf{r}_{mn}^{mov}\}$ and $\{\mathbf{r}_{mn}^{fix}\}$, $m = 1, \ldots, M_n$, are delineated in the moving- and fixed images, respectively. $\{\mathbf{r}_{mn}^{fix}\}$ denotes the ground-truth for the propagated "fixed ROI" $\mathbf{R}^{fix}$. These ROIs need not to be labelled consistently across image pairs, individual images pairs may have different types anatomical structures or regions as training ROI labels and may have different numbers of ROI pairs, i.e. in general, $M_1 \neq M_2 \ldots \neq M_n$.

The fixed ROIs in this work are represented by binary masks, indicating ground-truth class probabilities at each voxel $p(C_k|\mathbf{r}_{mn}^{fix})$, for a foreground $C_{k=1}$ and a background class $C_{k=0}$. Without loss of generality, the moving ROIs are also represented by binary masks, each as an input of the neural network that predicts the conditional class probabilities $p_\theta(C_k|\mathbf{i}_n^{fix}, \mathbf{i}_n^{mov}, \mathbf{r}_{mn}^{mov})$. Given $n^{th}$ image pair and $M_n$ associated ROI pairs, the negative log-likelihood leads to a weighted cross-entropy loss function: $J_n(\theta) = -\sum_{m=1}^{M_n} \sum_{k=1}^{2} p(C_k|\mathbf{r}_{mn}^{fix}) \log p_\theta(C_k|\mathbf{i}_n^{fix}, \mathbf{i}_n^{mov}, \mathbf{r}_{mn}^{mov}) w_k$, where the weighting parameter $w_k$ is the sample ratio between foreground and background voxels [11].

A typical image segmentation network, such as a 3D U-Net [12], can be adapted to take the input of an image pair and one of the moving ROI labels $(\mathbf{i}_n^{fix}, \mathbf{i}_n^{mov}, \mathbf{r}_{mn}^{mov})$. The previously-proposed two-stage sampling is adopted in a stochastic minibatch gradient descent optimisation, in which, image pairs are sampled first before sampling image-specific ROI labels. Thus, each minibatch has the same number of first-stage-sampled image pairs $(\mathbf{i}_n^{fix}, \mathbf{i}_n^{mov})$ and second-stage-sampled ROI pairs $(\mathbf{r}_{mn}^{mov}, \mathbf{r}_{mn}^{mov})$ and, collectively, they contribute to an unbiased estimator of the batch gradient [5]. During inference, given a new pair of images and a moving ROI, the trained network can predict where this ROI is propagated (or warped) to in the fixed image space.

### 2.3 Comparison to a DDF-Predicting Registration Network

We compared the proposed conditional segmentation network with a previously-proposed weakly-supervised registration network [5], because 1) it uses the same types of image and ROI data in training; and 2) it was proposed with a clinical aim for predicting ROIs, including the prostate gland, one or more image-visible lesions (potentially tumours) and surrounding organs, so these can be identified during ultrasound-guided interventional procedures [5, 8-10]. Once trained, the registration network does not need the moving ROI as input to predict a DDF for each image pair. Instead, it warps the ROI using the predicted DDF. The conditional segmentation network predicts a moved ROI directly, given the additional moving ROI. This difference is illustrated in Figs. 1 and 2. The details of both networks are summarised in Section 3.

**Registration Accuracy:** Two accuracy measures were computed in this study: Target registration error (TRE), defined as root-mean-square centroid distance, between the propagated moving ROIs and the ground-truth fixed ROIs, calculated over all ROI pairs for each test patient, and the Dice similarity coefficient (DSC) calculated between the pairs of ROIs representing the entire prostate. The training-independent TREs and DSCs are clinically informative in targeting the regions of surgical interest, such as prostate lesions, and in identifying vulnerable structures, such as rectum [9, 10]. They are reported based on the cross-validation experiments described in Section 3.

**Physically Plausible Correspondence Prediction for Out-of-Sample ROIs:** Predicting a new ROI at the inference stage could fail if this ROI is not within the ROI distribution represented by the training labels. However, it is reasonable to expect that a physically plausible mapping can be predicted on these novel landmarks using conditional segmentation without explicit deformation regularisation. This generalisability across different types of ROIs may be a result of potential anatomical, spatial and intensity correlations between these novel test ROIs and the training ROIs.

To test this generalisability, a set of *ad hoc* ROIs were selected if they do not have apparent representatives in the training data. For example, several patient-specific calcification clusters were found on unusual locations such as anterior regions of the prostate gland. The TREs on these ROIs are reported in addition to the overall results.

**Bias-Variance versus Training Data Size:** One of the potential advantages of avoiding deformation regularisation is to reduce the bias from the smoothness assumptions, such that more complicated correspondence can be learned from data, such as one-to-many or many-to-one mapping at voxel-level. Examples in this application include topological changes (presence of catheter in urethra and swelling during ablation) and high nonlinearity (between glandular zones and other structures).

To quantitatively investigate the bias for the two networks, we ran repeated experiments with bootstrap-sampled training/testing sets to decompose the variance due to random training data sampling and stochastic model training from the bias observed consistently across experiments. We propose two hypotheses, *Hypothesis A:* compared to the DDF-predicting network constrained by deformation regularisation (here, bending energy), the conditional segmentation would reduce the prediction bias, which is a component of the TRE; *Hypothesis B:* potential high-bias can limit generalisability in registration performance, represented by larger TREs on testing data, as training data increase. The Hypothesis B has an important practical value in informing the choice between these two types of networks, when training data size changes.

To test these hypotheses, we adopted a patient-level repeated cross-validation procedure [13] for both networks, by which, a set of training data sizes of interest is tested. The square of the bias and the variance on each ROI are then represented by the squared-distance $d_{bias}^2$ from the centre of the predicted centroids to the ground-truth and the average squared-distance $d_{var}^2$ to the centre from the centroids, respectively, over all samples estimated from the repeated cross-validation. This experiment does not take into account inter-training-data variability that will change as the training data size changes in cross-validation experiments, but it has been shown to be effective in estimating bias and variance of altering training data size [14], which was the concern in this study. The experiment details are described in Section 3.

## 3 Experiments

Without any initial alignment, a total of 115 pairs of T2-weighted MR and 3D transrectal ultrasound (TRUS) images from 80 prostate cancer patients who underwent TRUS-guided biopsy or therapy procedures were randomly sampled from clinical trial data (anonymised trial names and identifiers) for this study. Each patient may have multiple MR-TRUS image pairs according to the trial protocols. 3D TRUS volumes were reconstructed by rotational sagittal frames acquired by a bi-plane transrectal probe (Hitachi HI-VISION Preirus). All image volumes were normalised to zero-mean with unit-variance intensities after being resampled to 0.8×0.8×0.8 mm³ isotropic voxels. From these patients, a total of 910 pairs of corresponding anatomical ROIs were labelled and verified by second observers including consultant radiologists and senior imaging research fellows. Besides full gland segmentations for all cases, the ROIs defined landmarks including the apex and base of the prostate, the urethra, image-visible lesions, gland zonal separations, the vas deference and the seminal vesicles, and other patient-specific landmarks such as calcifications and fluid-filled cysts, with similar spatial and size distributions to those reported in the previous work [5].

Compared to the registration network architecture, only two changes were made to the original input and output layers to implement the conditional segmentation network: First, the additional single moving ROI label for each image pair, after being linearly-resampled to the fixed image size, is concatenated to the input layer with the image pair; second, instead of three displacement components, in *x*-, *y*- and *z*-channels, the conditional segmentation network outputs a single-channel logits layer, with a sigmoid function, to represent the foreground class probabilities in the fixed image space.

Compared to the registration network training, the conditional segmentation network training was found to be less sensitive to initialisation and learning rate as a result of not-predicting spatial transformations, required less memory without the 3D intensity resampler and did not need to tune the weighted deformation regularisation. Both networks had 32 initial channels and were trained with the same data using the Adam optimiser starting at a learning rate of $10^{-5}$. For brevity, readers are referred to the referenced publication and the published demonstration code [5] for additional details, which were kept unchanged in this work to enable comparison. The networks were implemented in TensorFlow™ with open-source code from NiftyNet [15]. Each of the conditional segmentation networks and registration networks was trained for 48 and 72 hours, respectively, both with a minibatch size of 2 using GeForce® GTX 1080Ti GPU cards with 11GB memory on a high-performance computing cluster.

As part of the patient-level repeated k-fold cross-validation procedure, the TREs of individual patients were calculated. The cross-validation was repeated ten times, for each of the four tested training patient sizes 40, 60, 70, and 75 with k = 2, 4, 8, and 16, respectively, with increasing yet varying numbers of image pairs (ranging from 40 to 110). This resulted in 600 networks trained in total (~36,000 GPU-hours) to compare the two types of networks. For each tested training size, the errors $d_{bias}^2$ and $d_{var}^2$ (described in Section 2.3), which represent the bias and variance in estimating ROI centroids, were estimated for individual ROIs from these ten samples.

## 4  Results

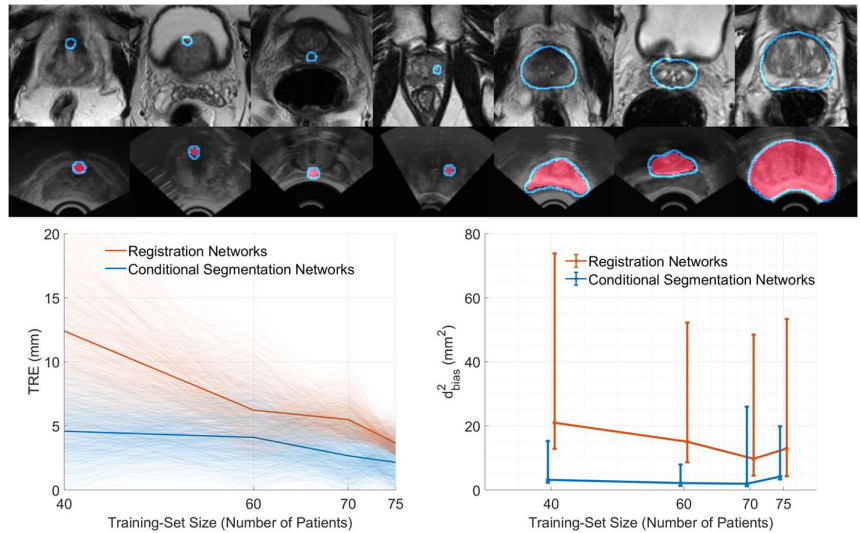

**Fig. 3.** Upper: example MR slices (1st row) with moving ROIs in blue contours, which are propagated to TRUS slices (2nd row) with the ground-truth ROIs in red areas; Lower: plots of mean TREs (left) and median $d_{bias}^2$ (right) versus training-set sizes. See the text for details.

From the 16-fold cross-validation, the median (25th - 75th percentiles) TRE and DSC are 2.1 (1.4-3.5) mm and 0.92 (0.90- 0.94), respectively, for the proposed conditional segmentation networks, and 3.2 (2.3-6.4) mm and 0.90 (0.87-0.92) for the DDF-predicting registration networks. A statistically significant ($p<0.001$) improvement of 34% in TREs was observed based on a paired Wilcoxon signed-rank test at a significance level $α=0.05$. TREs from the 70 manually selected *ad hoc* landmarks were also found to yield a lower median TRE of 2.8 (2.3-4.9) mm from the conditional segmentation, compared with 4.2 (3.0-7.8) mm using the registration network ($p<0.001$).

As training set sizes increase, the TREs were found to decrease for both networks with lower mean TREs being obtained from the conditional segmentation for all different training set sizes, as shown in Fig. 3. Furthermore, the estimated variance decreased with more training data, although detecting differences in variance between the two networks was difficult due to the small sample sizes (10 from the repeated cross-validation). The estimated bias from the conditional segmentation is considerably lower for all training set sizes as shown in the median (25$^{th}$ - 75$^{th}$ percentiles) $d_{bias}^2$ in Fig. 3. This suggests that 1) the prediction of spatial transformation may not be optimised for this task (here, a regularised DDF); 2) due to the high-and-non-decreasing bias observed from the registration network, the accuracy may not be improved by further increasing the training data; and 3) the lower TREs reported in this case can be attributed largely to the low-bias from the conditional segmentation network.

## 5      Conclusion

While prior knowledge, such as deformation regularisation, has proven useful for improving the generalisation ability of registration networks with limited training data [8], the bias-variance analysis presented in this paper revealed that this approach can also produce a high prediction bias, which may not be reduced by increasing the training data size. Thus, we have proposed a conditional segmentation paradigm for ROI propagation applications that avoids estimating a spatial transformation to overcome this issue. Using a supervised neural network, we have demonstrated significantly improved TREs in a real-world surgical application where clinically meaningful ROIs corresponding to those defined in MR images are predicted in prostate ultrasound images.

**Acknowledgement:** This work is supported by the Wellcome/EPSRC Centre for Interventional and Surgical Sciences (203145Z/16/Z) and TV is supported by a Medtronic / RAEng Research Chair (RCSRF1819/7/34), with additional support from CRUK (C28070/A19985), the Wellcome (203145Z/16/Z; 203148/Z/16/Z) and the EPSRC (EP/N026993/1, NS/A000050/1; NS/A000049/1).

## Reference

1. DeTone, D., Malisiewicz, T., & Rabinovich, A. (2016). Deep image homography estimation. arXiv preprint arXiv:1606.03798.
2. de Vos, B. D., et al., (2019). A deep learning framework for unsupervised affine and deformable image registration. *Medical image analysis*, 52, 128-143.
3. Eppenhof, K. A., et al., (2018). Deformable image registration using convolutional neural networks. *In Medical Imaging 2018: Image Processing*, 10574, 105740S
4. Rohé, M-M, et al., (2017). SVF-Net: Learning Deformable Image Registration Using Shape Matching. *MICCAI 2017*, LNCS 10433, 266-274.
5. Hu, Y., et al., (2018). Weakly-supervised convolutional neural networks for multimodal image registration. *Medical image analysis*, 49, 1-13.
6. Balakrishnan, G., et al., (2019). VoxelMorph: a learning framework for deformable medical image registration. *IEEE transactions on medical imaging*, in press.
7. Hering, A., et al., (2019). Enhancing Label-Driven Deep Deformable Image Registration with Local Distance Metrics for State-of-the-Art Cardiac Motion Tracking. *In Bildverarbeitung für die Medizin 2019*, 309-314

8. Hu, Y., et al., (2018). Adversarial Deformation Regularization for Training Image Registration Neural Networks. *MICCAI 2017*, LNCS 11070, 774-782
9. Siddiqui, M. M., et al., (2015). Comparison of MR/ultrasound fusion–guided biopsy with ultrasound-guided biopsy for the diagnosis of prostate cancer. Jama, 313(4), 390-397.
10. Valerio, M., et al., (2017). New and established technology in focal ablation of the prostate: a systematic review. *European urology*, 71(1), 17-34.
11. Lawrence, S., et al., (1998). Neural network classification and prior class probabilities. In Neural networks: tricks of the trade, 299-313
12. Çiçek, Ö., et al., (2016). 3D U-Net: learning dense volumetric segmentation from sparse annotation. *MICCAI 2016*, LNCS 9901, 424-432.
13. Webb, G. I. (2000). Multiboosting: A technique for combining boosting and wagging. *Machine learning*, 40(2), 159-196.
14. Webb, G. I., & Conilione, P. (2005). Estimating bias and variance from data. *Pre-publication manuscript* (http://users.monash.edu/~webb/Files/WebbConilione06.pdf).
15. Gibson, E. et al., (2018). NiftyNet: a deep-learning platform for medical imaging. *Computer Methods & Programs in Biomedicine*, 158, 113-122.